\documentclass[sigconf,natbib=false]{acmart}
\AtBeginDocument{%
 }

\copyrightyear{2026}
\acmYear{2026}
\setcopyright{cc}
\setcctype{by}
\acmConference[SVM '26]{4th International Workshop on Software Vulnerability Management }{April 12--18, 2026}{Rio de Janeiro, Brazil}
\acmBooktitle{4th International Workshop on Software Vulnerability Management (SVM '26), April 12--18, 2026, Rio de Janeiro, Brazil}
\acmPrice{}
\acmDOI{10.1145/3786165.3788441}
\acmISBN{979-8-4007-2397-1/2026/04}

\usepackage{algorithm}
\usepackage{algorithmic}

\usepackage{listings}
\usepackage{url}

\definecolor{diffadd}{rgb}{0, 0.6, 0} 
\definecolor{diffdel}{rgb}{0.8, 0, 0} 
\definecolor{codegray}{rgb}{0.5,0.5,0.5} 

\lstdefinestyle{c_diff}{
 language=C,
 basicstyle=\ttfamily\small, 
 keywordstyle=\color{blue},
 commentstyle=\color{green!60!black},
 stringstyle=\color{purple},
 numberstyle=\tiny\color{codegray}, 
 numbers=left, 
 stepnumber=1,
 numbersep=5pt,
 backgroundcolor=\color{white}, 
 showspaces=false, 
 showstringspaces=false, 
 showtabs=false, 
 frame=single, 
 rulecolor=\color{black},
 tabsize=2,
 captionpos=b, 
 breaklines=true, 
 breakatwhitespace=true, 
 title=\lstname, 
 escapeinside={\%*}{*)}, 
 morekeywords={xmlChar, xmlMalloc, size_t, SIZE_MAX, INT_MAX}, 
 moredelim=**[l][\color{diffadd}\bfseries]{@+}, 
 moredelim=**[l][\color{diffdel}\bfseries]{@-}, 
}

\definecolor{diffadd}{RGB}{0,128,0}
\definecolor{diffdel}{RGB}{180,0,0}
\definecolor{diffmeta}{RGB}{100,100,100}

\lstdefinelanguage{diff}{
  morecomment=[f][\color{diffmeta}]{diff --git},
  morecomment=[f][\color{diffmeta}]{index},
  morecomment=[f][\color{diffmeta}]{---},
  morecomment=[f][\color{diffmeta}]{+++},
  morecomment=[f][\color{diffmeta}]{@@},
  morecomment=[f][\color{diffadd}]{+},
  morecomment=[f][\color{diffdel}]{-},
}

\lstdefinestyle{githubdiff}{
  language=diff,
  basicstyle=\ttfamily\footnotesize,
  columns=fullflexible,
  keepspaces=true,
  showstringspaces=false,
  frame=single,
  breaklines=true,
  backgroundcolor=\color{white},
  captionpos=b, 
}

\RequirePackage[
 datamodel=acmdatamodel,
 style=acmnumeric,
 ]{biblatex}

\begin{document}

\title{Bridging Code Property Graphs and Language Models for Program Analysis}

\author{Ahmed Lekssays}
\affiliation{%
 \institution{Qatar Computing Research Institute}
 \city{Doha}
 \country{Qatar}
}
\email{alekssays@hbku.edu.qa}

\renewcommand{\shortauthors}{Ahmed Lekssays}

\begin{abstract}
Large Language Models (LLMs) face critical challenges when analyzing security vulnerabilities in real-world codebases: token limits prevent loading entire repositories, code embeddings fail to capture inter-procedural data flows, and LLMs struggle to generate complex static analysis queries. These limitations force existing approaches to operate on isolated code snippets, missing vulnerabilities that span multiple functions and files. We introduce \texttt{codebadger}, an open-source Model Context Protocol (MCP) server that integrates Joern's Code Property Graph (CPG) engine with LLMs. Rather than requiring LLMs to generate complex CPG queries, \texttt{codebadger} provides high-level tools for program slicing, taint tracking, data flow analysis, and semantic code navigation, enabling targeted exploration of large codebases without exhaustive file reading. We demonstrate its effectiveness through three use cases: (1) navigating an 8,000-method codebase to audit memory safety patterns, (2) discovering and exploiting a previously unreported buffer overflow in libtiff, and (3) generating a correct patch for an integer overflow vulnerability (CVE-2025-6021) in libxml2 on the first attempt. \texttt{codebadger} enables LLMs to reason about code semantically across entire repositories, supporting vulnerability discovery, patching, and program comprehension at scale.
\end{abstract}



\maketitle

\section{Introduction}

Large Language Models (LLMs) have advanced software vulnerability management (SVM), enabling automated tasks such as vulnerability detection \cite{nong2025appatch, jiao2025deepvulhunter, lin2025large, zhou2025large, liu2025improving, sovrano2025large}, patching \cite{bhandari2025generating, hu2025sok}, and repair \cite{kim2025logs, yu2025patchagent}. Recent benchmarks highlight \cite{yildiz2025benchmarking, realvulllm2025, zhu2025cve, haddad2025prompting, potter2025frontier} LLM capabilities through techniques like curriculum learning, agentic workflows, and multi-sample feedback. However, these approaches face limitations when applied to real-world codebases.

\textbf{The Challenge: LLMs Cannot Handle Large Codebases.} Security vulnerabilities rarely exist in isolation; they emerge from interactions across multiple functions, files, and modules. Consider a buffer overflow where the allocation size is determined in one function, the buffer is passed through several layers of abstraction, and the unsafe write occurs in a distant module. Detecting such vulnerabilities requires tracing data flows across the entire call graph, understanding control dependencies, and reasoning about buffer bounds at each step. Current LLM-based approaches fail at this task for three reasons:

\textit{First, token limits prevent loading entire repositories.} Modern codebases contain thousands of files and millions of lines of code. An 8,000-method repository like GGML would require hundreds of millions of tokens to load completely, far exceeding even extended context windows. Existing approaches resort to analyzing isolated functions or files, missing inter-procedural vulnerabilities \cite{lekssays2025llmxcpg}.

\textit{Second, code embeddings lack semantic depth.} While retrieval-augmented generation (RAG) systems can surface relevant code snippets via similarity search, embeddings capture syntactic patterns but fail to represent semantic relationships like ``function A's return value flows into function B's buffer write'' or ``this variable is validated in one branch but used unchecked in another.'' Such inter-procedural data flows and control dependencies are needed for vulnerability analysis yet remain invisible to embedding-based retrieval \cite{nong2025appatch}.

\textit{Third, LLMs struggle in generating complex static analysis queries.} Security analysts use static analysis tools with operations like backward slicing to isolate relevant code, taint tracking to trace untrusted data, and call graph traversal to understand propagation paths. While Code Property Graphs (CPGs) enable such analyses through query languages like CPGQL \cite{yamaguchi2014modeling}, LLMs struggle to generate correct queries for non-trivial analyses due to the scarcity of CPGQL in general-purpose training corpora. As CPGQL is a domain-specific language with complex traversal semantics, models frequently hallucinate API methods or produce syntactically invalid queries when attempting multi-hop traversals \cite{lekssays2025llmxcpg}. This fragility makes direct query generation unreliable for robust automation.

These limitations force a trade-off: either analyze small, isolated code snippets (missing cross-function vulnerabilities) or attempt to load large contexts (exceeding token budgets and degrading reasoning quality). Neither approach enables the analyst-like reasoning required for vulnerability management at scale.

\textbf{Our Solution: Providing LLMs with Analysis Tools.} We introduce \texttt{codebadger}, an open-source Model Context Protocol (MCP) \cite{mcp_spec} server that changes how LLMs interact with code analysis. Rather than requiring LLMs to read entire codebases or generate complex queries, \texttt{codebadger} integrates Joern's CPG engine with LLMs through high-level analysis tools. These tools (including program slicing, taint tracking, data flow analysis, call graph extraction, and bounds checking) abstract complex graph traversals into semantic operations that LLMs can invoke.

This design enables \textit{semantic navigation} rather than exhaustive reading. When analyzing an 8,000-method codebase for memory safety issues, an LLM agent using \texttt{codebadger} does not read all 8,000 methods. Instead, it calls \texttt{get\_\allowbreak codebase\_\allowbreak summary} to understand scale, \texttt{find\_\allowbreak taint\_\allowbreak sources} to locate 54 allocation sites, \texttt{find\_\allowbreak taint\_\allowbreak sinks} to identify 300+ dangerous operations, and \texttt{find\_\allowbreak taint\_\allowbreak flows} to trace specific data flows, consuming only the tokens needed for relevant code snippets. This mirrors how human analysts work: they do not read linearly but navigate semantically, following paths through the call graph and dependency chains.

By providing pre-implemented analyses, \texttt{codebadger} eliminates the need for LLMs to master CPGQL syntax or graph traversal algorithms. An LLM reasoning about a potential buffer overflow can call \texttt{get\_\allowbreak program\_\allowbreak slice} with a suspicious call site, receiving a minimal code snippet containing all relevant data and control dependencies. It can trace variable origins with \texttt{get\_\allowbreak data\_\allowbreak dependencies}, verify bounds checking with \texttt{find\_\allowbreak bounds\_\allowbreak checks}, and explore propagation paths with \texttt{get\_\allowbreak call\_\allowbreak graph}, all without generating CPGQL.

\textbf{Contributions.} We demonstrate \texttt{codebadger}'s effectiveness through three use cases. In code comprehension, an LLM agent audited the GGML library's memory safety across 8,667 methods and 198,006 calls, identifying issues like unbounded \texttt{alloca} usage and integer overflows in allocation size calculations. In vulnerability discovery, the agent found and validated an unreported buffer overflow in libtiff by tracing \texttt{col\_offset} through pointer arithmetic, generating an ASAN-confirmed exploit. In vulnerability patching, it analyzed CVE-2025-6021 in libxml2, using data flow analysis and program slicing to generate a correct patch on the first attempt that closely matched the maintainers' fix.
These results show that \texttt{codebadger} enables LLMs to perform context-aware reasoning across entire repositories, supporting tasks such as vulnerability discovery, patching, bug repair, and program analysis in agentic setups. The tool is available at \url{https://github.com/lekssays/codebadger}.

\section{Background}

\subsection{Code Property Graphs}

Code Property Graphs (CPGs) combine multiple code representations into a single graph structure \cite{yamaguchi2014modeling}. CPGs integrate Abstract Syntax Trees (ASTs) for syntactic structure, Control Flow Graphs (CFGs) for execution paths, and Program Dependence Graphs (PDGs) for data and control dependencies. This unified representation enables queries over code properties using graph traversal languages such as the CPG Query Language (CPGQL).

Tools like Joern implement CPG generation and querying for various programming languages, including C/C++, Java, and Python. CPGs are valuable in vulnerability detection by allowing the identification of patterns such as taint flows from untrusted sources to sensitive sinks \cite{lekssays2025llmxcpg}. By traversing these graphs, analysts can detect issues like buffer overflows or injection vulnerabilities without executing the code.

\subsection{Model Context Protocol}

The Model Context Protocol (MCP) is a framework that enables Large Language Models (LLMs) to interact with external tools and services through a consistent interface. MCP exposes functions as callable tools over HTTP, allowing AI assistants to augment their capabilities with specialized operations such as data retrieval or analysis.

In software analysis, MCP facilitates the integration of LLMs with static analysis engines. It supports session-based management, asynchronous processing, and caching to handle complex queries. This protocol is useful for providing LLMs with contextual information from large datasets or tools, overcoming limitations like token constraints and enabling agentic workflows where models can dynamically invoke tools during reasoning.

\section{Related Work}

Recent advancements in Large Language Models (LLMs) have impacted software vulnerability management, with studies exploring their applications in detection, patching, and repair. This section reviews key works in these areas and positions \texttt{codebadger} as a tool that enhances LLM capabilities through static analysis.

\subsection{LLM-based Vulnerability Detection}

Several studies have evaluated LLMs for vulnerability detection, revealing limitations in handling complex, real-world codebases. Yildiz et al. \cite{yildiz2025benchmarking} benchmark LLMs and LLM-based agents on vulnerability detection tasks, showing good results on function-level benchmarks but performance degradation in inter-procedural scenarios.

Other works focus on improving detection through different techniques. Liu et al. propose a curriculum preference learning framework to improve LLM vulnerability detection by synthesizing reasoning data. Chen et al. explore reinforcement learning-based fine-tuning for LLMs. Nong et al. \cite{nong2025quality} investigate using LLMs to improve vulnerability data quality by filtering low-quality instances.

These approaches primarily rely on direct prompting or fine-tuning, often struggling with semantic context in large repositories. \texttt{codebadger} addresses this by providing LLMs with CPG-based tools for precise slicing and taint analysis, enabling integration into detection pipelines like those in \cite{lekssays2025llmxcpg}, where CPG-guided LLMs achieve 15-40\% F1-score improvements.

\subsection{LLM-based Vulnerability Patching and Repair}

Research on LLM-driven patching and repair has gained traction, with models automating fix generation \cite{bhandari2025generating, yu2025patchagent, hu2025sok}. Zhang et al. introduce Vul-R2, a reasoning LLM for automated repair, using domain-aware reasoning to decompose patching into stages. Kim et al. \cite{kim2025logs} present SAN2PATCH, which uses multi-stage LLM reasoning for vulnerability patching based on logs and tree-of-thought analysis.

Evaluation frameworks like VulnRepairEval by Ma et al. assess LLM patching using exploit-based metrics, revealing inconsistencies in repair quality across languages. Hu et al. propose multi-sample self-reward feedback to improve LLM patching accuracy.

While these methods show promise, they often lack deep static analysis integration. \texttt{codebadger}'s tools, such as call graph extraction and program slicing, can augment these repair workflows by supplying targeted code contexts.

\texttt{codebadger} is designed for integration into agentic workflows. By exposing CPG tools via MCP, it allows LLMs to dynamically invoke analyses like taint flow tracking or reachability checks during multi-step reasoning. In vulnerability discovery, agents can chain slicing with detection prompts; in patching, they can extract dependencies before generating fixes; in program analysis, they can browse call graphs for bug repair. Unlike pure LLM approaches, \texttt{codebadger} provides semantic fidelity, reducing hallucinations and enabling security tasks across repositories.

\section{Methodology and Implementation}

This section describes the methodological framework of \texttt{codebadger} and its implementation. Our approach integrates Joern's Code Property Graph (CPG) analysis with the Model Context Protocol (MCP) to provide Large Language Models (LLMs) with tools for static code analysis. Rather than serving as an interface for executing raw CPG Query Language (CPGQL) queries, \texttt{codebadger} provides higher-level abstractions that encapsulate complex graph traversals and analyses. This enables LLMs to perform reasoning similar to human security analysts, facilitating tasks such as inter-procedural vulnerability detection, patching, and program comprehension. The methodology is grounded in program analysis principles, drawing from techniques like backward slicing and taint propagation \cite{yamaguchi2014modeling, lekssays2025llmxcpg}, while addressing LLM limitations in query generation and context management.

\subsection{Architecture}

\begin{figure*}[t]
\centering
\includegraphics[width=6in]{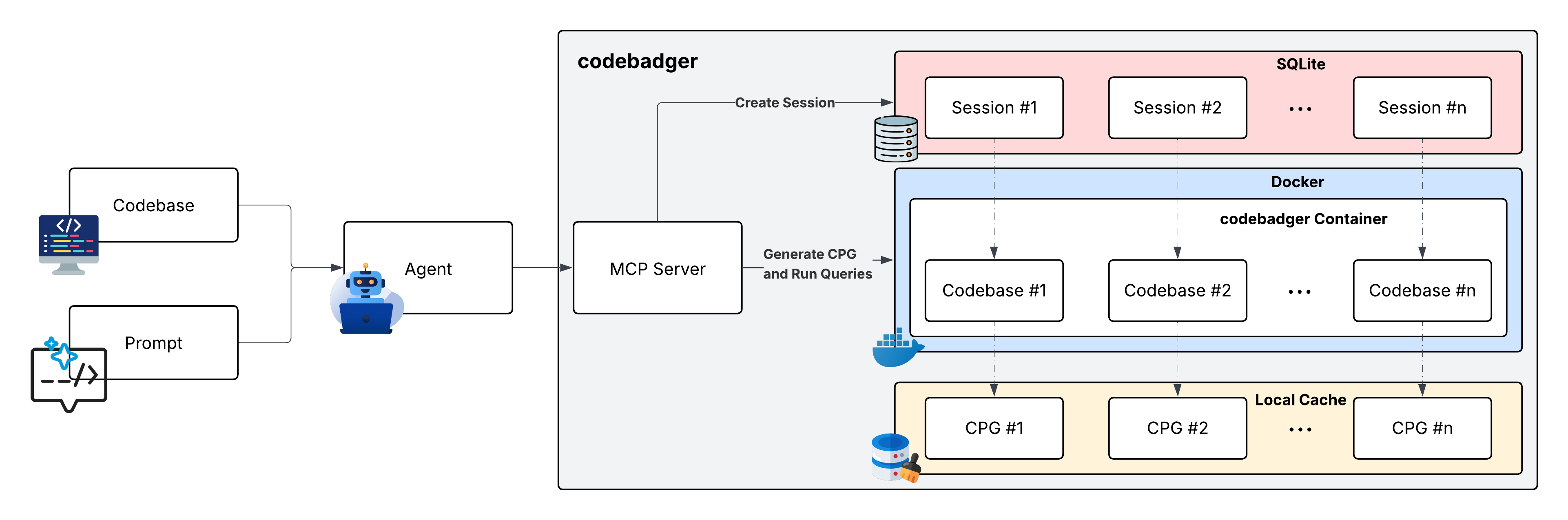}
\caption{Architecture of \texttt{codebadger}.}
\label{fig:architecture}
\end{figure*}

As illustrated in Figure \ref{fig:architecture}, \texttt{codebadger} acts as an intermediary service between an LLM agent and the Joern static analysis engine. The workflow is session-based and managed via MCP over HTTP.

The process begins when the LLM agent calls \texttt{create\_\allowbreak cpg\_\allowbreak session}, passing a repository. The \texttt{codebadger} server invokes Joern to parse the code and generate a CPG, which is cached. The server returns a \texttt{session\_id} to the LLM. The agent can then use this ID to make subsequent tool calls, such as \texttt{get\_\allowbreak program\_\allowbreak slice} or \texttt{get\_\allowbreak data\_\allowbreak dependencies}. The MCP server translates these high-level functions into specific CPGQL queries, executes them via Joern, and returns the structured results (e.g., code snippets, JSON-formatted data flows) to the LLM. This abstracts the complexity of graph traversal, allowing the LLM to focus on analysis.

\subsection{Framework}

The core of \texttt{codebadger}'s methodology includes security-focused analyses for vulnerability detection, supplemented by code browsing capabilities for navigation. These components support agentic workflows, where LLMs can iteratively invoke tools to build contextual understanding without exhausting token budgets on extraneous code.

Code browsing tools enable syntactic and semantic exploration, mitigating the challenges of large codebases. For instance, call graph construction tools build directed graphs of method invocations up to a specified depth, using CPG traversals to map caller-callee relationships. Similarly, method source extraction and code snippet retrieval tools pull targeted code segments, while literal search tools employ pattern matching to locate hardcoded values, aiding in the identification of potential misconfigurations.

The primary emphasis lies in the security analysis tools. These implement algorithms for data flow analysis, taint tracking, and program slicing, leveraging Joern's data-flow engine to perform vulnerability-oriented computations. For example, taint analysis for detecting flows from untrusted sources to sensitive sinks is realized through dedicated taint flow identification and path listing mechanisms. Algorithm \ref{alg:taint} outlines the taint propagation process, adapted from standard data-flow techniques \cite{lekssays2025llmxcpg}.

\begin{algorithm}[t]
\caption{Taint Flow Analysis Algorithm}
\label{alg:taint}
\begin{algorithmic}[1]
\REQUIRE CPG $G$, Sources $S$, Sinks $K$, MaxPaths $M$
\ENSURE Set of taint paths $P$
\STATE $P \gets \emptyset$
\STATE Identify taint sources: $sources \gets \{ n \in G \mid n \in S \}$ 
\COMMENT{Using CPGQL for source patterns}
\STATE Identify taint sinks: $sinks \gets \{ n \in G \mid n \in K \}$ 
\COMMENT{Using CPGQL for sink patterns}
\FOR{each $src \in sources$}
  \STATE Compute forward data flows: $flows \gets$ traverse PDG from $src$ tracking variable propagations
  \FOR{each path $path \in flows$ that reaches a $sink \in sinks$}
    \IF{$|P| < M$}
      \STATE $P \gets P \cup \{path\}$ 
      \COMMENT{Collect paths with data/control dependencies}
    \ENDIF
  \ENDFOR
\ENDFOR
\RETURN $P$ as JSON-structured paths
\end{algorithmic}
\end{algorithm}

This algorithm begins by querying the CPG for predefined source and sink nodes (e.g., input functions like \texttt{read} as sources and execution points like \texttt{exec} as sinks). It then performs forward traversal along the Program Dependence Graph (PDG) edges, propagating taint labels while capturing data and control dependencies. To manage computational complexity, the number of paths is capped, preventing exponential growth in large graphs. Argument flow detection extends this by focusing on parameter propagations, useful for detecting injection vulnerabilities.

Program slicing employs backward slicing to isolate vulnerability-relevant code segments, reducing codebase size by up to 90\% as demonstrated in empirical studies \cite{lekssays2025llmxcpg}. Algorithm \ref{alg:slice} details this process.
Starting from a criterion point (e.g., a potentially vulnerable function call), the algorithm iteratively collects upstream dependencies via PDG and CFG traversals, ensuring the slice preserves semantic equivalence for the targeted behavior. This focused extraction enhances LLM reasoning by eliminating noise.

\begin{algorithm}[t]
\caption{Backward Program Slicing Algorithm}
\label{alg:slice}
\begin{algorithmic}[1]
\REQUIRE CPG $G$, Criterion point $c$ (line or call)
\ENSURE Slice $slice$ containing dependencies
\STATE $slice \gets \{c\}$
\STATE $worklist \gets \{c\}$
\WHILE{$worklist \neq \emptyset$}
  \STATE $node \gets$ pop from $worklist$
  \STATE Add data dependencies: $deps \gets$ traverse PDG backward from $node$ for variables/expressions affecting $node$
  \STATE Add control dependencies: $ctrls \gets$ traverse CFG backward for conditionals governing $node$'s execution
  \STATE $slice \gets slice \cup deps \cup ctrls$
  \STATE $worklist \gets worklist \cup (deps \cup ctrls) \setminus slice$
  \COMMENT{Avoid revisiting}
\ENDWHILE
\STATE Reconstruct code: $code \gets$ serialize $slice$ into source code snippet
\RETURN $code$
\end{algorithmic}
\end{algorithm}

Method reachability verification utilizes call graph traversals to confirm inter-method paths, supporting analyses of propagation across modules.

\subsection{Implementation Details}

\texttt{codebadger} is implemented as a FastMCP \cite{fastmcp} (v2.12.4) server using Streamable HTTP for concurrent interactions. It runs on a configurable host and port, with YAML-based settings for logging and parameters. Each session creates an isolated analysis context by initializing a CPG from a local path or git repository. The CPG is stored in a dedicated environment for efficient graph queries, which are handled asynchronously. Sessions run in Docker containers using a pre-built Joern image for CPG generation and queries. This ensures isolation, security, and reproducibility. Redis manages session state and caching. CPGs are cached by source hash and language to avoid recomputation, asynchronous jobs are tracked in queues, and results are stored for polling. With configurable concurrency, the system scales efficiently for high-throughput analysis.

\section{Use Cases}

In this section, we explore several use cases that demonstrate how \texttt{codebadger} enables Large Language Models (LLMs) to perform program analysis tasks. To evaluate our approach, we address three research questions linked to the challenges identified in Section 1:

\begin{itemize}
    \item \textbf{RQ1 (Scale):} Can \texttt{codebadger} enable analysis of codebases that exceed LLM context windows?
    \item \textbf{RQ2 (Semantics):} Does the semantic context provided by CPG tools enable the discovery of inter-procedural vulnerabilities missed by snippet-based analysis?
    \item \textbf{RQ3 (Usability):} Can \texttt{codebadger} abstract complex static analysis effectively, enabling LLMs to perform expert-level tasks without generating raw queries?
\end{itemize}

By leveraging the tools provided by \texttt{codebadger}, LLMs can perform activities ranging from code comprehension to vulnerability detection and patching. We used Claude Sonnet 4.5 as the LLM for these examples, but \texttt{codebadger} is compatible with any agentic framework or command-line tool that supports MCP. We note that the knowledge cutoff date of Claude Sonnet 4.5 is January 2025, so it does not have prior knowledge of vulnerabilities or codebases discussed hereafter. We use GitHub Copilot Agent as the orchestration framework to manage the LLM interactions with \texttt{codebadger}.

\subsection{Code Browsing and Comprehension}

Beyond vulnerability-specific tasks, \texttt{codebadger} enables LLMs to perform code exploration and security auditing across large codebases. We demonstrate this through a security evaluation of memory accesses in the GGML library, a tensor computation framework comprising 1,151 files and approximately 246k lines of code (LOC), with over 8,000 methods.

\textbf{Setup.} The agent was tasked with evaluating memory access security patterns across the entire GGML codebase. It received access to \texttt{codebadger} tools like \texttt{create\_\allowbreak cpg\_\allowbreak session} and \texttt{get\_\allowbreak codebase\_\allowbreak summary}. The agent established a CPG session for the local GGML repository.

\textbf{LLM Reasoning and Tool Usage.} The agent initiated exploration by calling \texttt{get\_\allowbreak codebase\_\allowbreak summary} to understand the scale: 8,667 methods across 198,006 function calls. Rather than reading files sequentially, it employed a targeted strategy using \texttt{find\_\allowbreak taint\_\allowbreak sources} to locate 54 memory allocation sites, and \texttt{find\_\allowbreak taint\_\allowbreak sinks} to identify 300+ potentially dangerous operations.

The agent then used \texttt{list\_\allowbreak methods} to find functions performing bulk memory operations, narrowing focus to security-critical code paths. For each suspicious pattern, it invoked \texttt{get\_\allowbreak code\_\allowbreak snippet} to retrieve context. For instance, in \texttt{src/ggml-alloc.c}, it identified \texttt{malloc} calls and used \texttt{find\_\allowbreak taint\_\allowbreak flows} to trace whether \texttt{hash\_set.size} originated from validated input.

Using \texttt{run\_\allowbreak cpgql\_\allowbreak query}, the agent executed custom CPGQL queries to find allocations with arithmetic in size parameters. It discovered 20+ instances requiring validation. The agent also queried for \texttt{snprintf} calls to assess format string safety, finding 100+ instances with hardcoded buffer sizes, most handled safely but some lacking external input validation.

\textbf{Results and Insights.} The agent identified three issues: (1) use of \texttt{alloca} without size bounds, risking stack overflow; (2) integer overflows in complex allocation size calculations; (3) unchecked \texttt{op\_\allowbreak params} pointer arithmetic causing potential out-of-bounds reads. It also noted 12 issues including missing \texttt{realloc} return value checks and inadequate buffer size validation in quantization functions. Positive findings included consistent use of \texttt{sizeof} and size-limited functions like \texttt{snprintf}.

\begin{lstlisting}[style=c_diff, caption={Integer Overflow Risk in GGML Allocation Size Calculation}, label={lst:ggml_overflow}]
// src/ggml-backend.cpp:1624
// Note: Multiple multiplications might overflow before allocation
// If (sched->n_backends * sched->n_copies) is large enough, the result wraps around, allocating a small buffer.
malloc(sched->hash_set.size * sched->n_backends * sched->n_copies * sizeof(struct ggml_tensor *))
\end{lstlisting}

This case study demonstrates \texttt{codebadger}'s utility for code comprehension at scale. By providing query primitives for source/sink identification and taint analysis, the agent avoided exhaustive file reading (which would exceed token limits for an 8,000-method codebase) and instead navigated semantically, focusing on memory safety hotspots. The tools enabled a structured audit workflow: summarize $\rightarrow$ locate sources/sinks $\rightarrow$ trace flows $\rightarrow$ validate patterns, mirroring human analyst practices but automated through LLM reasoning over CPG abstractions.

\subsection{Vulnerability Detection}

We demonstrate \texttt{codebadger}'s effectiveness in discovering previously unreported vulnerabilities through a case study involving \texttt{libtiff} (828 files, 127k LOC), specifically the \texttt{tif\_getimage.c} file. The analysis was performed by an LLM agent without prior knowledge of the vulnerability.

\textbf{Setup.} The agent was tasked with analyzing \texttt{tif\_getimage.c} for buffer underflow vulnerabilities. It was provided with access to the following \texttt{codebadger} tools: \texttt{create\_\allowbreak cpg\_\allowbreak session}, \texttt{list\_\allowbreak methods}, \texttt{get\_\allowbreak method\_\allowbreak source}, \texttt{find\_\allowbreak taint\_\allowbreak sources}, \texttt{find\_\allowbreak taint\allowbreak \_sinks}, \texttt{get\_\allowbreak data\_\allowbreak dependencies}, \texttt{find\_\allowbreak bounds\_\allowbreak checks}, and \texttt{run\_\allowbreak cpgql\_\allowbreak query}. The agent established a CPG session for the \texttt{libtiff} repository at the vulnerable commit\footnote{\url{https://gitlab.com/libtiff/libtiff/-/commit/14da830be4bf4d6d3daafcbd892426aa6c164242}}.

\textbf{LLM Reasoning and Tool Usage.} The agent initiated analysis by calling \texttt{list\_\allowbreak methods} to identify tile and strip reading functions in \texttt{tif\_getimage.c}, discovering critical methods including \texttt{gtTileContig} and \texttt{gtStripContig}. It then retrieved their source code using \texttt{get\_\allowbreak method\_\allowbreak source} to examine buffer access patterns. The agent identified suspicious pointer arithmetic involving \texttt{pos} calculations, where \texttt{fromskew} and \texttt{col\_offset} values were used without validation.

To trace the origin of these potentially unsafe values, the agent employed \texttt{list\_\allowbreak calls} to locate user-controlled input sources. It discovered that \texttt{col\_offset} and \texttt{row\_offset} were set via the \texttt{TIFFRG\allowbreak BAImage} structure without bounds checking. Using \texttt{get\_\allowbreak data\_\allowbreak dependencies} with backward analysis on the \texttt{pos} variable at line 801, the agent traced data flow from \texttt{col\_offset} through \texttt{fromskew} calculations, confirming that user-controlled values directly influenced buffer pointer arithmetic.

The agent then called \texttt{find\_\allowbreak bounds\_\allowbreak checks} targeting the buffer access locations, revealing that bounds validation occurred \emph{after} the buffer access rather than before, creating a time-of-check-time-of-use vulnerability pattern. Finally, using \texttt{run\_\allowbreak cpgql\_\allowbreak query}, it confirmed that no validation existed to ensure \texttt{col\_offset < tile\_width} or \texttt{row\_offset < tile\_height} before the offset calculations. This analysis enabled the agent to identify a buffer overflow vulnerability where maliciously large offset values could cause \texttt{buf + pos} to point outside allocated memory bounds.

Listing~\ref{lst:vuln_code} shows the vulnerable code identified by the agent. The critical issue occurs in the \texttt{gtStripContig()} function where \texttt{img->col\_offset} is used in pointer arithmetic without validation, allowing \texttt{pos} to overflow when \texttt{col\_offset} exceeds \texttt{imagewidth}.

\begin{lstlisting}[style=c_diff, caption={Vulnerable Code in gtStripContig() - No Validation of col\_offset}, label={lst:vuln_code}, escapechar=@]
static int gtStripContig(TIFFRGBAImage *img, uint32_t *raster, 
 uint32_t w, uint32_t h) {
 TIFF *tif = img->tif;
 tileContigRoutine put = img->put.contig;
 uint32_t imagewidth = img->width;
 tmsize_t scanline, pos;
 //... initialization code...
 
 for (row = 0; row < img->height; row += nrow) {
 //... strip reading code...
 
 @\color{red}// VULNERABLE: No validation that col\_offset is within image bounds@
 pos = ((row + img->row_offset) % rowsperstrip) * scanline +
 ((tmsize_t)img->col_offset * img->samplesperpixel);
 
 tmsize_t roffset = (tmsize_t)y * w;
 @\color{red}// HEAP-BUFFER-OVERFLOW when col\_offset > imagewidth@
 (*put)(img, raster + roffset, 0, y, w, nrow, 
 fromskew, toskew, buf + pos);
 }
}
\end{lstlisting}

\textbf{Results and Insights.} This use case addresses \textbf{RQ2} and \textbf{RQ3}. The vulnerability was inherently inter-procedural, requiring the tracking of \texttt{col\_offset} from the \texttt{TIFFRGBAImage} structure across method boundaries to the computation in \texttt{gtStripContig}. Standard RAG techniques retrieving individual functions would likely miss the struct definition and initialization context required to identify the missing check. Moreover, the agent successfully performed complex taint tracking and bounds checking using high-level tools (\texttt{find\_taint\_sources}, \texttt{get\_data\_dependencies}), avoiding the need to construct intricate CPGQL queries (RQ3) while maintaining semantic precision (RQ2).

\subsection{Exploit Generation}

Following vulnerability discovery, the agent leveraged \texttt{codebadger} to construct a proof-of-concept (PoC) exploit and validate the finding using AddressSanitizer (ASAN).

\textbf{LLM Reasoning and Tool Usage.} The agent analyzed the vulnerable code path to determine exploitable conditions. Using \texttt{get\_\allowbreak program\_\allowbreak slice} on the identified call in \texttt{gtStripContig}, it obtained a minimal code context showing the complete calculation: \texttt{pos = ((row + img->row\_offset) \% rowsperstrip) * scanline + ((tmsize\_t)img->col\_offset * img->samplesperpixel)}. The agent reasoned that setting \texttt{col\_offset} to exceed the image width would produce a \texttt{pos} value beyond the allocated buffer size.

The agent then synthesized a PoC in C that created malicious TIFF files with carefully crafted offset values. It tested multiple attack vectors: (1) \texttt{col\_offset = 356} with \texttt{width = 256} for strip-based images, and (2) \texttt{col\_offset = 512} with \texttt{tile\_width = 256} for tile-based images. To validate the vulnerability, the agent compiled both \texttt{libtiff} and the PoC with ASAN instrumentation flags. Upon execution, ASAN detected a heap-buffer-overflow at the predicted location:

\begin{verbatim}
==1009465==ERROR: AddressSanitizer: heap-buffer-overflow
READ of size 1 at 0x52b00000d200
 #0 putRGBcontig8bittile tif_getimage.c:1701
 #1 gtStripContig tif_getimage.c:1157
 #2 TIFFRGBAImageGet tif_getimage.c:603
\end{verbatim}

This confirmed the vulnerability was exploitable with attacker-controlled TIFF files containing malicious \texttt{col\_offset} values.

\textbf{Verification of Unreported Status.} To determine if the vulnerability was previously known, the agent discovered commit \texttt{68169402}\footnote{\url{https://gitlab.com/libtiff/libtiff/-/commit/681694024846f543fe7d4821074b813cd9dccdfa}} titled "Improve TIFFReadRGBAImage and avoid buffer overflows," dated June 25, 2025. Examining this commit, the agent found it introduced validation checks rejecting \texttt{col\_offset >= img->width} and \texttt{row\_offset >= img->height}, directly addressing the discovered vulnerability. 

Listing~\ref{lst:patched_code} shows the fix applied by the maintainers. The patch adds validation to ensure \texttt{col\_offset} is within valid bounds before performing buffer arithmetic, and calculates a safe width (\texttt{wmin}) to prevent out-of-bounds access.

\begin{lstlisting}[style=c_diff, caption={Patched Code in gtStripContig() - Added Validation of col\_offset}, label={lst:patched_code}, escapechar=@]
static int gtStripContig(TIFFRGBAImage *img, uint32_t *raster, 
 uint32_t w, uint32_t h) {
 TIFF *tif = img->tif;
 tileContigRoutine put = img->put.contig;
 uint32_t imagewidth = img->width;
 tmsize_t scanline, pos;
 @\color{purple}uint32\_t wmin;@
 
 @\color{purple}// FIX: Validate col\_offset before use@
 @\color{purple}if (0 <= img->col\_offset \&\& @
 @\color{purple} (uint32\_t)img->col\_offset < imagewidth) \{@
 @\color{purple}wmin = TIFFmin(w, imagewidth - img->col\_offset);@
 @\color{purple}\} else \{@
 @\color{purple}TIFFErrorExtR(tif, TIFFFileName(tif),@
 @\color{purple}"Error in gtStripContig: column offset \%d "@
 @\color{purple}"exceeds image width \%d",@
 @\color{purple}img->col\_offset, imagewidth);@
 @\color{purple}return 0;@
 @\color{purple}\}@
 
 for (row = 0; row < img->height; row += nrow) {
 //... strip reading code...
 
 pos = ((row + img->row_offset) % rowsperstrip) * scanline +
 ((tmsize_t)img->col_offset * img->samplesperpixel);
 
 tmsize_t roffset = (tmsize_t)y * w;
 @\color{purple}// Now uses wmin instead of w to limit copying to valid data@
 (*put)(img, raster + roffset, 0, y, @\color{purple}wmin@, nrow, 
 fromskew, toskew, buf + pos);
 }
}
\end{lstlisting}

The agent then switched to the master branch, rebuilt with ASAN, and re-ran the PoC, which now failed with the error message: \texttt{"Error in gtStripContig: column offset 356 exceeds image width 256"}. This confirmed the vulnerability was previously unreported until fixed in that commit. No CVE identifier was assigned to this issue in the available git history or security advisories.

\textbf{Results and Insights.} This result reinforces \textbf{RQ2} (Semantics). Generating a functional exploit requires precise semantic understanding of data flow arithmetic—specifically, how an integer input leads to a calculated offset. The ability to calculate the exact values (e.g., \texttt{356}, \texttt{512}) required for the PoC demonstrates that the LLM possessed a deep, semantic model of the program's execution state, far beyond the capabilities of simple pattern matching or syntactic completion.

\subsection{Vulnerability Patching}

To illustrate how an LLM can utilize \texttt{codebadger}, we present an analysis of a recorded interaction where an LLM agent (Claude Sonnet 4.5) investigates a known vulnerability in the \texttt{xmlBuildQName} function of the \texttt{libxml2} library (1,574 files, 335k LOC) that is vulnerable to an integer overflow that causes a buffer overflow (CVE-2025-6021)\footnote{\url{https://gitlab.gnome.org/GNOME/libxml2/-/issues/926}}.

\textbf{Setup.} We provided the agent with the original issue submitted to the library maintainers and asked it to patch it. The issue included a PoC and ASAN output. The agent did not have access to any other external tools or internet access. The agent was given access to the following \texttt{codebadger} tools: \texttt{create\_\allowbreak cpg\_\allowbreak session}, \texttt{list\_\allowbreak methods}, \texttt{get\_\allowbreak method\_\allowbreak source}, \texttt{list\_\allowbreak calls}, \texttt{get\_\allowbreak data\_\allowbreak dependencies}, and \texttt{get\_\allowbreak program\_\allowbreak slice}. The agent was provided with the repository URL of \texttt{libxml2} and instructed to create a CPG session for it.

\textbf{LLM Reasoning and Tool Usage.} The LLM initiated the analysis by establishing a context for the investigation through \texttt{codebadger}. Recognizing the need for static analysis, it first called \texttt{create\_\allowbreak cpg\_\allowbreak session} to load or generate the Code Property Graph (CPG) for the \texttt{libxml2} C codebase. The server confirmed loading a cached CPG, providing the necessary \texttt{session\_id} for subsequent operations. This foundational step allowed the LLM to access the structured representation of the code needed for deeper analysis.

With the CPG session active, the LLM began focusing on the specific area of interest. It employed the \texttt{list\_\allowbreak methods} tool with the pattern \texttt{"xmlBuildQName"} to pinpoint the target function within the potentially large codebase. This mimics an analyst narrowing their scope, efficiently locating the function in \texttt{tree.c}. To understand the function's internal logic, the LLM then requested its source code using \texttt{get\_\allowbreak method\_\allowbreak source}. This provided the exact implementation details necessary for identifying potential security flaws.

The LLM then shifted its focus towards identifying potentially dangerous operations, a common step in security reviews. It reasoned that \texttt{memcpy} calls are common sources of buffer overflows if not handled carefully. Using \texttt{list\_\allowbreak calls}, it specifically searched for \texttt{memcpy} calls originating from within \texttt{xmlBuildQName}. This targeted query revealed two instances at lines 189 and 191, highlighting them as critical points for further investigation.

To understand the context of the second \texttt{memcpy} call, the LLM investigated the origins of the variables controlling the copy size: \texttt{lenn} and \texttt{lenp}. It performed backward data flow analysis using two calls to \texttt{get\_\allowbreak data\_\allowbreak dependencies}. This revealed that both variables were directly derived from \texttt{strlen} calls on potentially input-controlled strings (\texttt{ncname} and \texttt{prefix}). This step confirms that the size parameters of the dangerous sink (\texttt{memcpy}) are influenced by external inputs.

Finally, to gain a comprehensive understanding of the factors influencing the risky \texttt{memcpy}, the LLM requested a \texttt{get\_\allowbreak program\_\allowbreak slice}. This tool call provided a contextual snippet including both data dependencies (like the \texttt{strlen} assignments to \texttt{lenn} and \texttt{lenp}) and control dependencies (like the buffer allocation logic based on \texttt{memory == NULL} or size checks). Analyzing this slice allowed the LLM to identify the core vulnerability: the calculation \texttt{lenn + lenp + 2} could potentially lead to an integer overflow, resulting in an undersized buffer allocation via \texttt{xmlMalloc}, followed by buffer overflows in the subsequent \texttt{memcpy} calls. The sequence of tool calls enabled the LLM to gather sufficient evidence and context to accurately diagnose the integer overflow and propose a correct patch.

This interaction demonstrates how \texttt{codebadger} facilitates a structured, security-focused reasoning process for an LLM. By providing tools for targeted navigation (\texttt{list\_\allowbreak methods}, \texttt{list\_\allowbreak calls}), context gathering (\texttt{get\_\allowbreak method\_\allowbreak source}, \texttt{get\_\allowbreak program\_\allowbreak slice}), and data flow tracing (\texttt{get\_\allowbreak data\_\allowbreak dependencies}), the LLM could effectively mimic an analyst's workflow to identify, understand, and ultimately address the vulnerability.

\textbf{Proposed Patch.} The LLM proposed the following patch based on its analysis. The diff highlights changes to use \texttt{size\_t} for lengths, add explicit checks for integer overflow before calculating the total required size, and use the calculated \texttt{total\_len} consistently. We note that this patch was generated in the first attempt without any human intervention.

 
 
 
 

\begin{lstlisting}[style=githubdiff,
  caption={GitHub-style Diff for the \texttt{xmlBuildQName} Fix (CVE-2025-6021)},
  label={lst:patch_diff}]
diff --git a/tree.c b/tree.c
index e2357131..7a82ee23 100644
--- a/tree.c
+++ b/tree.c
@@ -164,10 +164,10 @@ xmlGetParameterEntityFromDtd(...)
- int lenn, lenp;
+ size_t lenn, lenp;
  xmlChar *ret;

- if (ncname == NULL) return(NULL);
+ if ((ncname == NULL) || (len < 0)) return(NULL);
  if (prefix == NULL) return((xmlChar *) ncname);

@@ -178,8 +178,10 @@ xmlBuildQName(...)
  lenn = strlen((char *) ncname);
  lenp = strlen((char *) prefix);
+ if (lenn >= SIZE_MAX - lenp - 1)
+     return(NULL);

- if ((memory == NULL) || (len < lenn + lenp + 2)) {
+ if ((memory == NULL) || ((size_t) len < lenn + lenp + 2)) {
\end{lstlisting}

After applying the patch, the LLM verified that the integer overflow check prevents unsafe buffer allocations, mitigating the vulnerability. Interestingly, the proposed patch closely mirrors the actual fix implemented by the \texttt{libxml2} maintainers \footnote{\url{https://gitlab.gnome.org/GNOME/libxml2/-/commit/ad346c9a249c4b380bf73c460ad3e81135c5d781}}, demonstrating the LLM's capability to reason about vulnerabilities using \texttt{codebadger}'s static analysis tools. This use case shows how \texttt{codebadger} enables LLMs to conduct static analyses and identify vulnerabilities in real-world codebases.

\textbf{Results and Insights.} This case answers \textbf{RQ2} and \textbf{RQ3} in the context of remediation. The successful patch relied on backward slicing to understand the integer overflow root cause, a semantic property invisible to key-value lookups (RQ2). Furthermore, the agent generated the correct fix by invoking abstract tools (e.g., \texttt{get\_program\_slice}), effectively bypassing the difficulty of writing complex graph traversal queries to isolate the overflow condition (RQ3).

\section{Discussion}
\texttt{codebadger} bridges LLMs and static code analysis for vulnerability management by providing pre-implemented tools that abstract complex Code Property Graph (CPG) operations. This enables LLMs to reason over large codebases without raw query generation pitfalls or context overload. Our use case demonstrates how \texttt{codebadger} facilitates vulnerability patching through targeted analyses like data dependency tracing and program slicing, successfully identifying and remediating CVE-2025-6021 in libxml2. Beyond patching, \texttt{codebadger} supports vulnerability discovery and generic program analysis. For discovery, tools identifying taint sources (e.g., \texttt{read}, \texttt{getenv}) and sinks (e.g., \texttt{exec}, \texttt{memcpy}), combined with path listing, enable LLMs to detect security-relevant data flows. Code browsing capabilities support selective navigation for uncovering inter-procedural vulnerabilities. For generic analysis, these tools enable dependency mapping and literal searching for tasks like refactoring or debugging.

\texttt{codebadger} has its own limitations. First, Joern's CPG generation creates overhead for very large repositories—massive codebases like the Linux kernel require substantial RAM that can be prohibitive in resource-constrained environments. Second, despite high-level abstractions, LLMs occasionally require direct CPGQL queries via \texttt{run\_\allowbreak cpgql\_\allowbreak query}, where Claude Sonnet 4.5 still fails sometimes, consistent with prior work \cite{lekssays2025llmxcpg}. Third, \texttt{codebadger} inherits Joern's limitations in handling runtime vulnerabilities like race conditions. Finally, effectiveness depends on the LLM's tool-chaining ability, which varies across models. While our qualitative case studies demonstrate practical utility in identifying real-world vulnerabilities, large-scale quantitative evaluation against benchmark datasets remains critical future work to rigorously assess generalizability.

\subsection{Ethical Considerations}

In adherence to ACM policies on responsible disclosure, vulnerabilities discovered during \texttt{codebadger} development were handled ethically. Two zero-day vulnerabilities—one in a widely used image parsing library and another in a popular machine learning framework—were identified using our taint analysis tools. These were promptly reported to the respective maintainers prior to any public disclosure, allowing time for patches. No human subjects were involved in this research. We acknowledge the use of LLMs in polishing the final version of this manuscript and to debug complex CPGQL queries.

\section{Conclusion}

In conclusion, \texttt{codebadger} provides LLMs with static analysis capabilities through its MCP-integrated tools, enabling vulnerability management in complex codebases. By abstracting CPGQL complexities into accessible functions for slicing, taint tracking, and browsing, it facilitates agentic workflows similar to security analysts, as evidenced by its role in analyzing and patching real-world vulnerabilities. The open-source tool, available at \url{https://github.com/lekssays/codebadger}, contributes to AI-driven SVM and invites community extensions. For future work, we plan to investigate at scale the use of taint analysis tools by LLMs, automating processes to monitor and optimize LLM behaviors in large-scale vulnerability detection pipelines. This could include integrating dynamic analysis or expanding support for emerging languages and systems.

\section*{Acknowledgements}
We thank the anonymous reviewers for their insightful feedback and constructive suggestions, which significantly improved the quality of this work.

\printbibliography

\end{document}